\documentclass[twocolumn,aps,pra,floatfix,longbibliography]{revtex4-2}

\usepackage{amsmath}
\usepackage{txfonts}
\usepackage{microtype}
\usepackage{graphicx}
\usepackage{color}
\usepackage{ulem}

\DeclareMathOperator{\arcsinh}{arcsinh}

\begin{document}

\title{Tunneling ionization in ultrashort laser pulses:  edge-effect and remedy }

\author{M. Klaiber}\email{klaiber@mpi-hd.mpg.de}
\author{Q. Z. Lv}\email{qingzheng.lyu@mpi-hd.mpg.de}
\author{K. Z. Hatsagortsyan}
\author{C. H. Keitel}
\affiliation{Max-Planck-Institut f\"ur Kernphysik, Saupfercheckweg 1, 69117 Heidelberg, Germany}

\begin{abstract}

Tunneling ionization of an atom in ultrashort laser pulses is considered.  When the driving laser pulse is switched-on and -off with a steep slope, the photoelectron momentum distribution (PMD) shows an edge-effect because of the photoelectron diffraction by the time-slit of the pulse. The trivial diffraction pattern of the edge effect consisting of fast oscillations in the PMD disguises in the deep nonadiabatic regime the physically more interesting features in the spectrum which originate from the photoelectron dynamics. We point out the precise conditions how to avoid this scenario  experimentally and if unavoidable in theory we put forward an efficient method to remove the edge-effect in the PMD. This allows to highlight the nonadiabatic dynamical features of the PMD, which is indispensable for their further investigation in complex computationally demanding scenarios. The method is firstly demonstrated on a one-dimensional problem, and further applied in three-dimensions for the attoclock. The method is validated by a comparison of analytical results via the strong-field approximation with numerical solutions  of the time-dependent Schr\"odinger equation.

\end{abstract}

\date{\today}

\maketitle

\section{Introduction}

Modern state-of-the-art laser techniques allow for full control over the wave form of a laser pulse, and in particular,  the generation of few-cycle strong laser pulses \cite{Brabec_2000,Paulus_2003,Goulielmakis_2004,Piccoli_2021}. Such few-cycle laser pulses of sufficient strength are an efficient tool in attoscience \cite{Corkum_2007,Ivanov_2005R,Krausz_2009}. They have been employed for the generation of isolated attosecond pulses via high-order harmonic generation (HHG) \cite{Drescher_2001,Agostini_2004,Calegari_2016,Popmintchev_2012}, for molecular imaging and laser induced electron diffraction \cite{Niikura_2003,Itatani2004,Blaga_2012,Wolter_2016}, as well as  for the time-resolved study of strong-field phenomena, such as nonsequential double ionization \cite{Moshammer_2000,Bhardwaj_2001,Rudenko_2004,Wang_2009,Kuebel_2016,Chen_2017,Chen_2019,Liu_2021} and dissociative ionization  \cite{Kling_2006,Xu_2017}. The theoretical description of strong-field phenomena in few-cycle pulses within the strong field approximation (SFA) is outlined in Ref.~\cite{Milosevic_2006}.

In ultrashort laser pulses an abrupt switch-on and -off of the laser pulse can induce a diffraction effect of the photoelectrons by the time slit of the pulse due to the pulse edges, the so-called edge-effect. The edge-effect is exhibited as oscillations in the photoelectron momentum distribution (PMD), additionally to  the dynamical features of PMD, and it disappears in the case of a smooth laser pulse. The edge-effect distorts the most important dynamical physical signal in strong-field ionization and for this reason one tries to avoid or separate it. The distortion is especially conspicuous at low laser intensities when the ionization signal is weak, but just in this deeply nonadiabatic regime the dynamical features of PMD are nontrivial. We underline that there are observed unexplained features in  PMD in elliptically polarized laser fields in the weak field regime \cite{Landsman_2014o,Ivanov_2014}, and the edge-effect hinders  their analysis.

In an experiment the role of edge-effects could be diminished using more and more smoother laser pulses.
In a theoretical description via  numerical solution of time-dependent Schr\"odinger equation (TDSE), as well as within  SFA, different forms of laser pulses with a  smooth switch-on and -off are employed. The most simple description of a short $N$-cycle laser pulse is via a $\cos^2$-envelope: $f(t)=\cos^2 (\omega t/N)$, with the laser frequency $\omega$, see e.g.  \cite{Milosevic_2006,Martiny_2008}.
Smoother pulses are obtained via $\cos^n$-envelopes with $n=4$ or larger (in this case one needs to take into account  the change of the effective frequency of the laser field). A better description is obtained with the use of a Gaussian pulse with a  long tail \cite{Patchkovskii_2016}, which however requires rather time consuming computationally expensive  calculations.

In this paper we put forward a simple method to separate the edge-effect and single-out the PMD dynamical signal in SFA calculations as well as in the numerical solution of TDSE, while using laser pulses with no-smooth switching. The method (U-contour method) mimics the saddle-point integration, however, without  explicit finding and classification of \textit{all} relevant saddle-points for the given PMD. We demonstrate the method in a one-dimensional (1D) model of tunneling ionization in  half-cycle  pulses of $\cos^2$ and truncated-Gaussian form, and confirm its accuracy in comparison with the numerical TDSE solution. Finally, we apply the method in a 3D example of the  attoclock. The U-contour method in 3D has a clear advantage  with respect to the saddle-point integration, as the latter would require the calculation of a large data set of saddle-points.

The structure of the paper is the following. In Sec.~\ref{sec-SFA} the SFA model is introduced,and the edge-effect is described. The conditions for the appearance of the effect are discussed in Sec.~\ref{sec-conditions}. The U-contour method for separation of the edge-effect is introduced in Sec.~\ref{sec-U-contour}, and its performance is tested in comparison with numerical solutions of TDSE. The application of the U-contour method for the analysis of the edge-effect in 3D case of attoclock is presented in Sec.~\ref{sec-attoclock}, and the conclusion is given in Sec.~\ref{sec-Conclusion}.
\begin{figure}
   \begin{center}
    \includegraphics[width=0.4\textwidth]{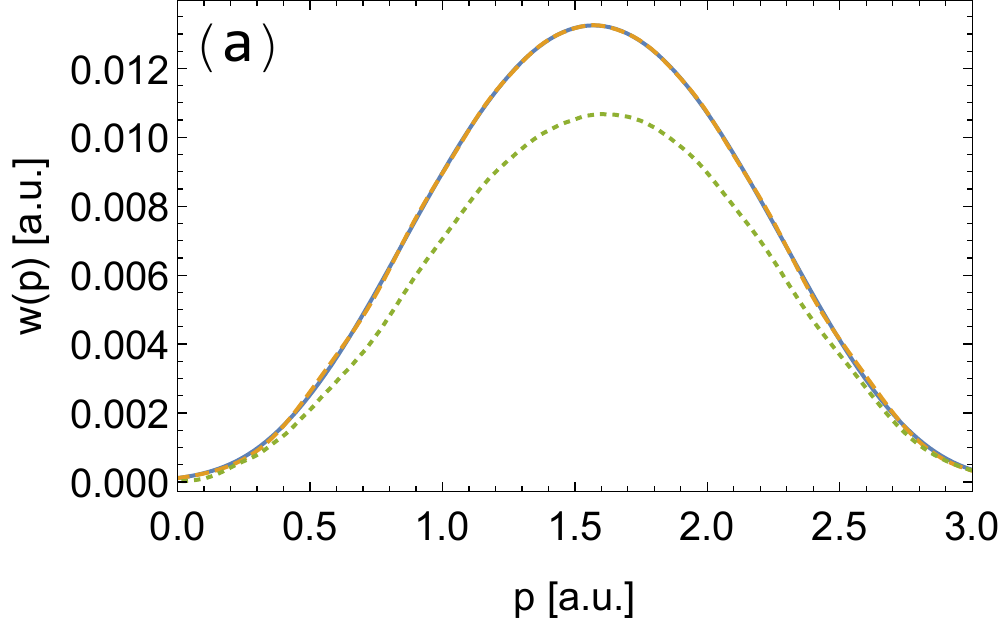}
   \includegraphics[width=0.4\textwidth]{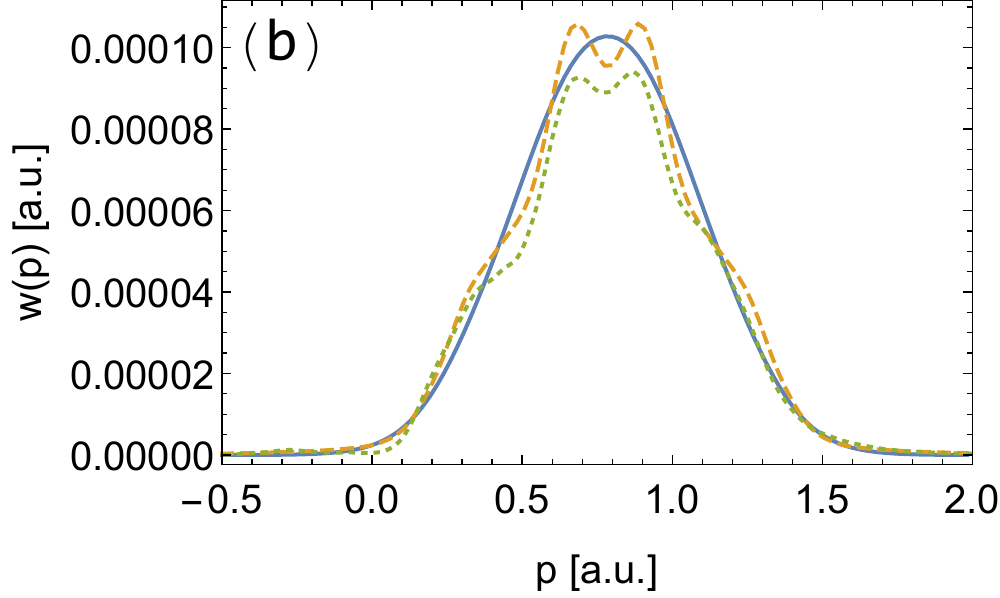}
   \includegraphics[width=0.4\textwidth]{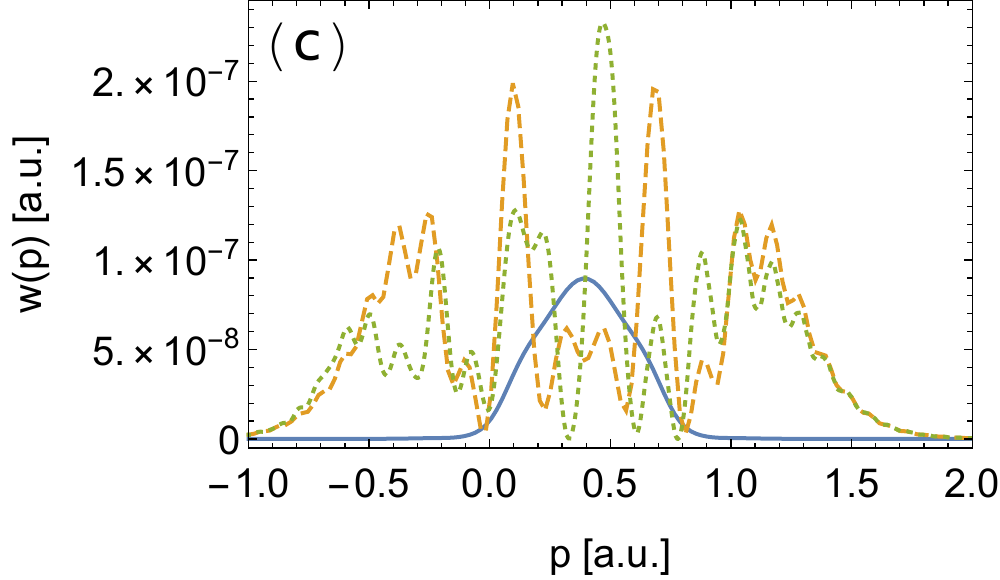}
            \caption{PMD  with  a field $E(t)=-E_0 \cos^2(\omega t)$: (a) $E_0=0.1$, (b) $E_0=0.05$, (c) $E_0=0.025$, $\omega=0.05$ a.u., $\kappa=1$ a.u., the field is truncated at $\omega t_i=-\pi/2$ and $\omega t_f=\pi/2 $, (orange, dashed) via SFA with the edge-effect, (blue, solid) via SFA with the edge-effect subtracted, (dotted, green) via numerical solution of TDSE. }
       \label{sin}
    \end{center}
  \end{figure}

  \begin{figure*}
   \begin{center}
     \includegraphics[width=0.4\textwidth]{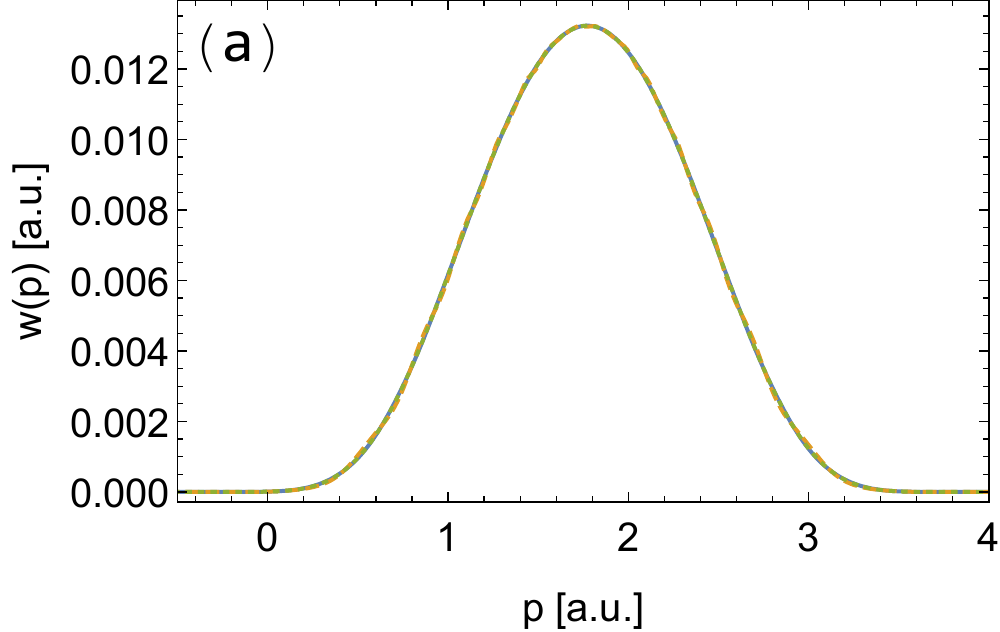}
     \includegraphics[width=0.4\textwidth]{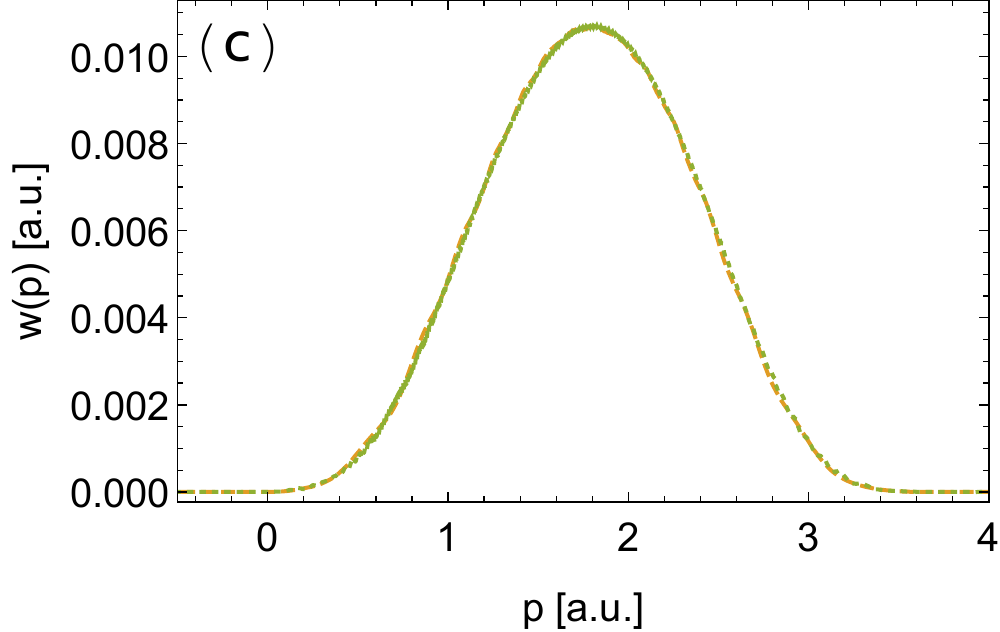}
     \includegraphics[width=0.4\textwidth]{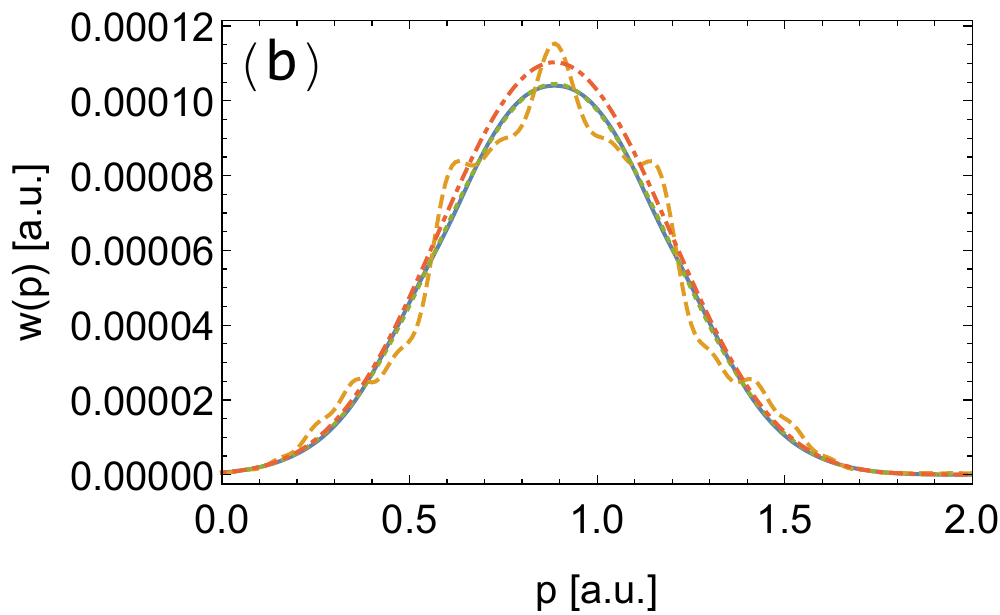}
     \includegraphics[width=0.4\textwidth]{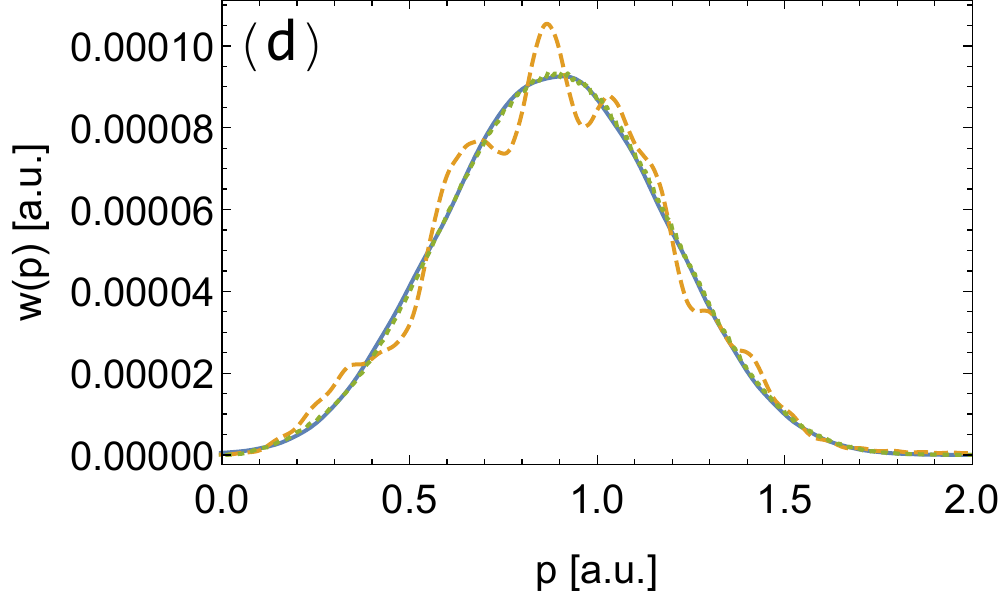}
            \caption{PMD with a field $E(t)=-E_0 \exp[-(\omega t)^2]$: (left column) via first-order SFA Eq.~(\ref{direct}); (right column) via numerical solution of TDSE; (a,c) $E_0=0.1$, (b,d)) $E_0=0.05$; (dashed, orange) short pulse with the truncation points of the Gaussian at $\omega t_i=-2$ and $\omega t_f=2$, (dotted, green) long pulse with $\omega t_i=-8$ and $\omega t_f=8$, (blue, solid) the edge-effect subtracted, (red, dot-dashed) in (b) is SFA calculation with SPA; $\omega=0.05$ a.u., $\kappa=1$ a.u.}
       \label{Gaussian}
    \end{center}
  \end{figure*}
  \begin{figure*}
   \begin{center}
     \includegraphics[width=0.4\textwidth]{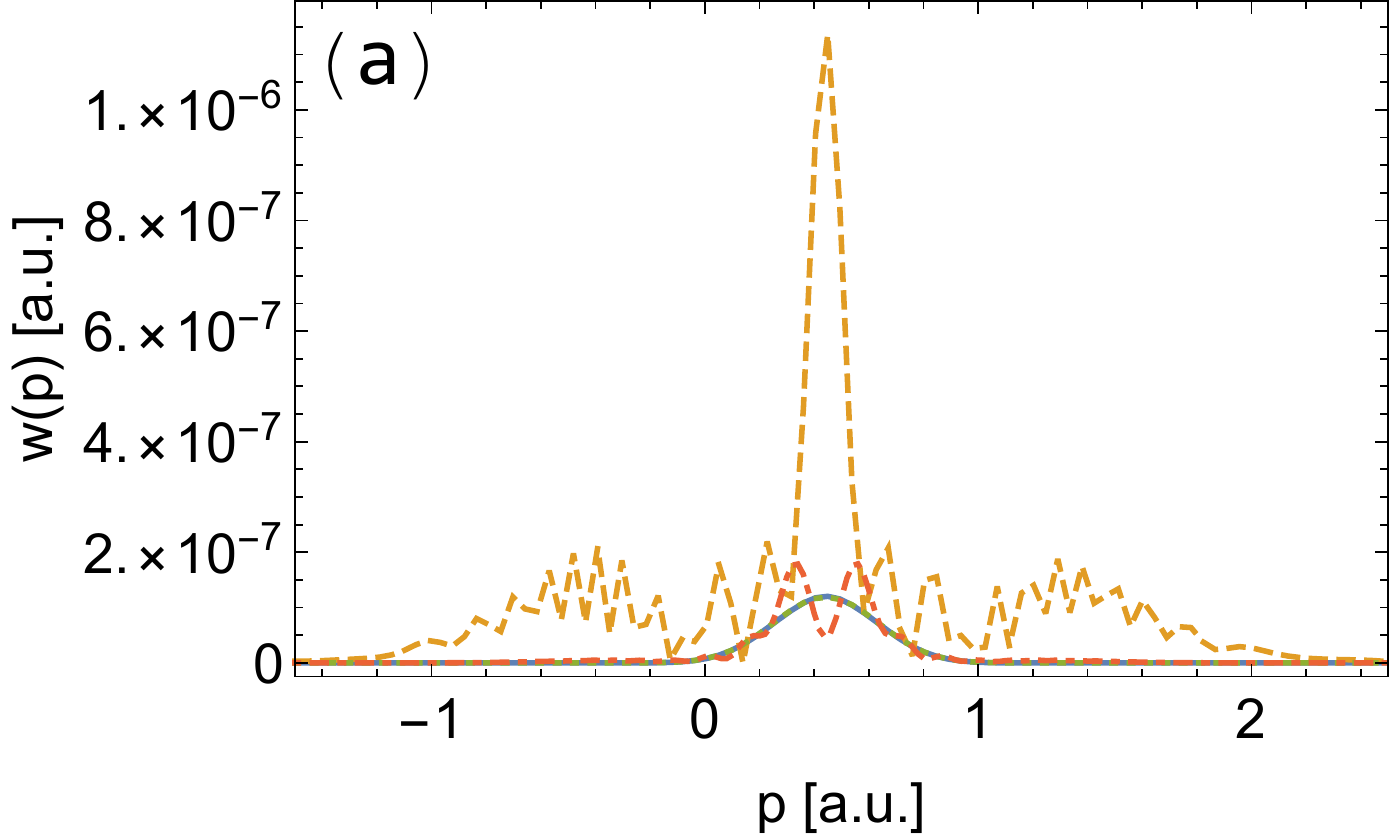}
   \includegraphics[width=0.4\textwidth]{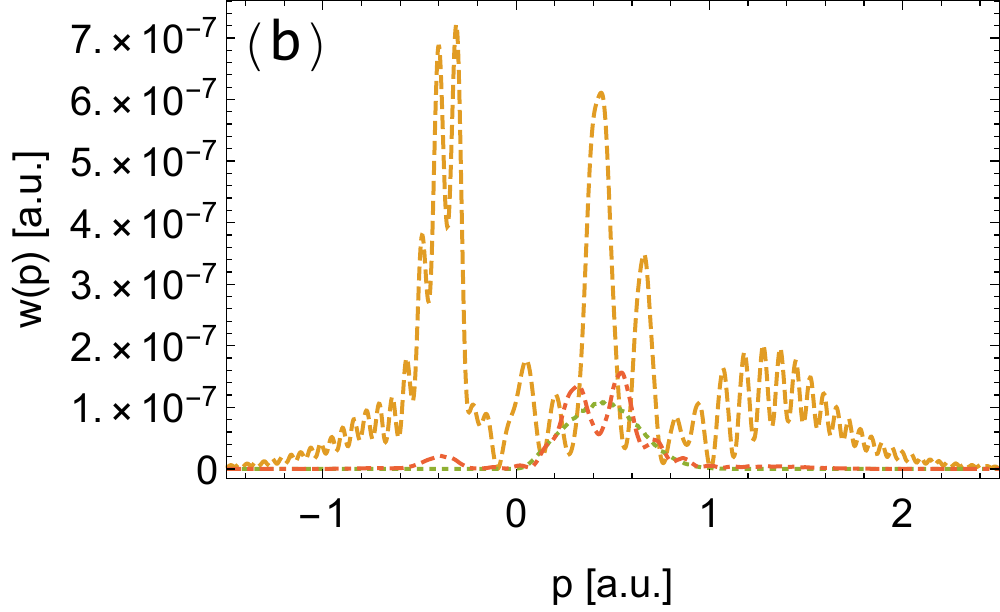}
                  \caption{PMD with a field $E(t)=-E_0 \exp[-(\omega t)^2]$, $E_0=0.025$: (a) via first-order SFA, (b) via numerical solution of TDSE; (dashed, orange) short pulse with the truncation points of the Gaussian at $\omega t_i=-2$ and $\omega t_f=2$, (dotted, green) long pulse with $\omega t_i=-8$ and $\omega t_f=8$, (blue, solid) SFA with the edge-effect subtracted, (dot-dashed, red) via Patchkowsii's truncated Gaussian \cite{Patchkovskii_2016} with parameters $\omega t_1=2$, and $\omega t_2=2.5$; $\omega=0.05$ a.u., $\kappa=1$ a.u.}
                      \label{fig3}
    \end{center}
  \end{figure*}
  \begin{figure}
   \begin{center}
 \includegraphics[width=0.45\textwidth]{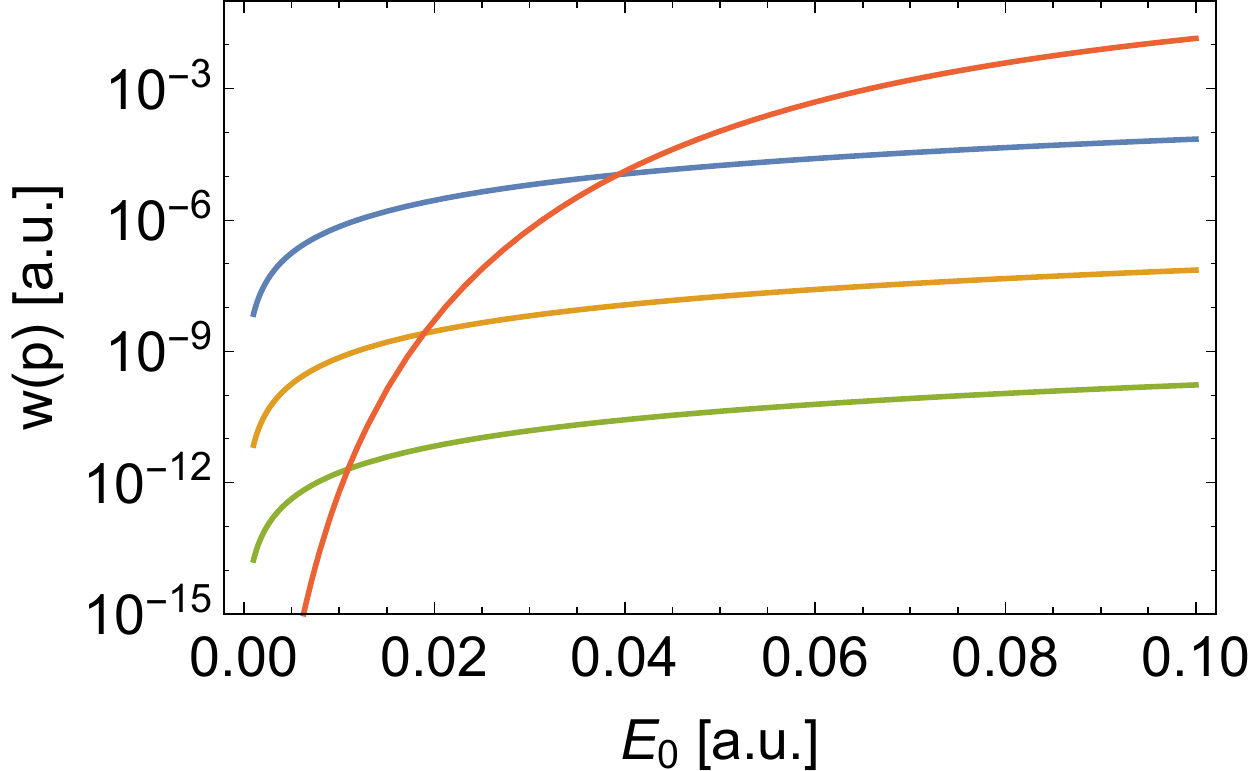}
           \caption{Determination of the threshold field strength in $\cos^n$-pulses, below which the edge-effect contaminates the strong-field ionization PMD. The edge-effect is induced by the high-energy component of the field with $\Omega \gtrsim I_p$, the probability of photoionization $W_\Omega$ via such an one high-energy photon absorption: (blue) in $\cos^2$-pulse, (orange) in $\cos^4$-pulse, (green) in $\cos^6$-pulse; (red) the  PPT-probability $W_{SFI}$. The edge-effect will be visible below the field value when $W_\Omega=W_{SFI}$.}
       \label{sinn}
    \end{center}
  \end{figure}
\begin{figure}
   \begin{center}
 \includegraphics[width=0.4\textwidth]{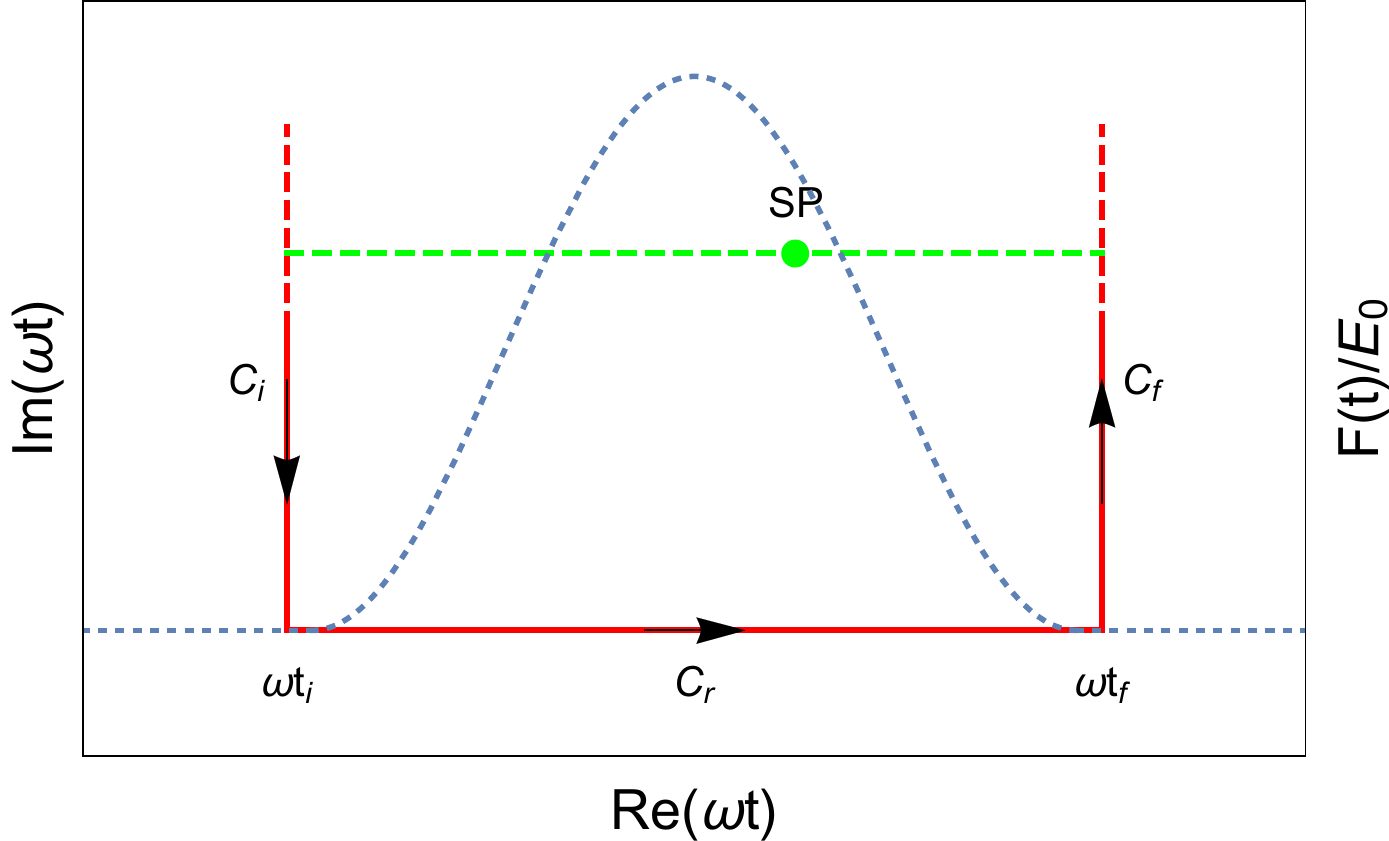}
              \caption{ Contours of the time-integration in Eqs.~(\ref{m_p}) and (\ref{eq:pmd_num_fin}): (red, horizontal) the original contour along the real time-axis from the pulse outset $t_i$ to the end $t_f$, (green, dashed) the saddle-point contour, (red, solid) the proposed U-contour to remove the pulse edge-effect. The laser pulse form is illustrated via the blue, dashed line.  }
                   \label{contours}
    \end{center}
  \end{figure}
\begin{figure}
   \begin{center}
   \includegraphics[width=0.4\textwidth]{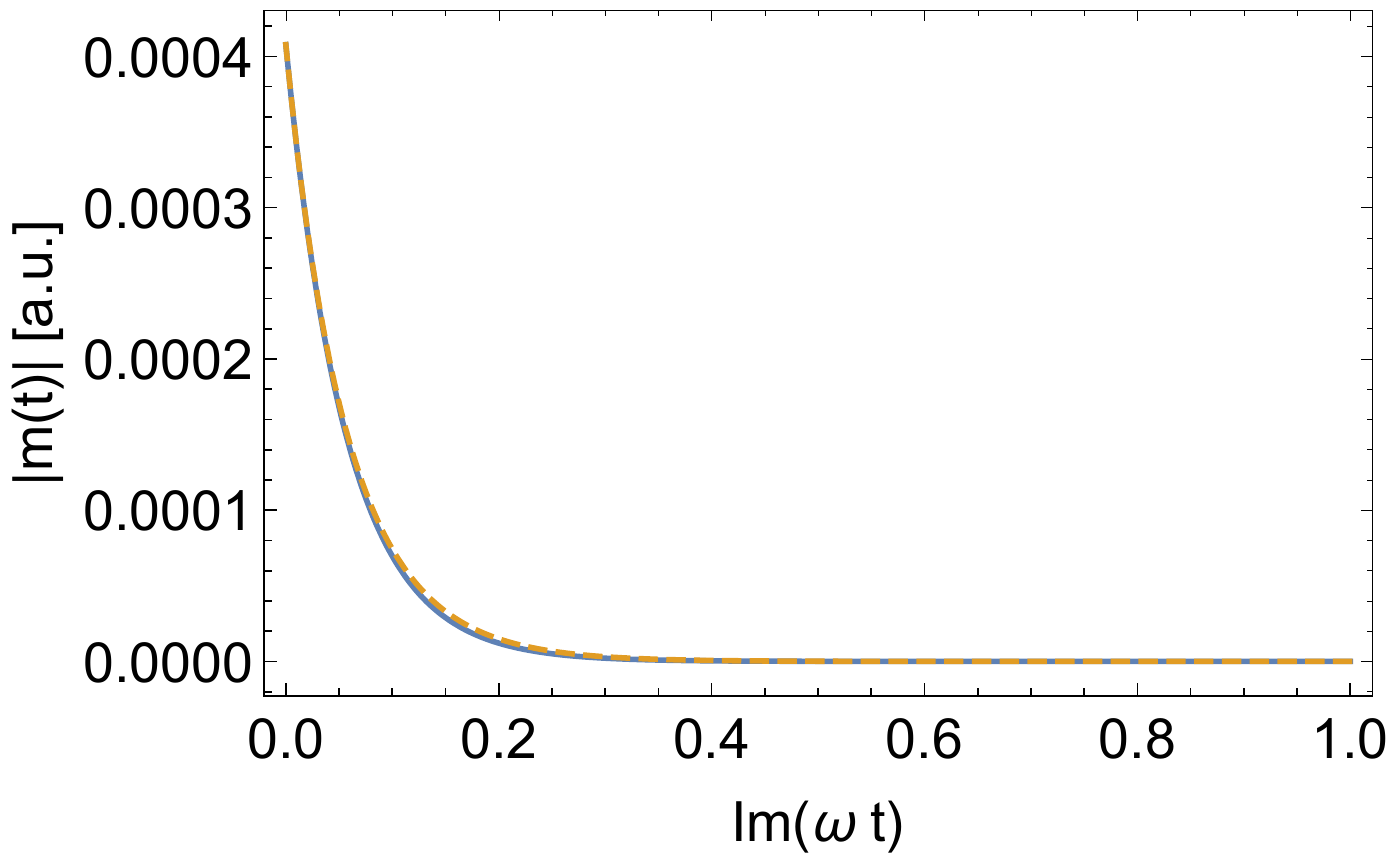}
            \caption{ The amplitude of the integrand along the vertical contour, exactly (blue, solid), and approximated by the expansion of Eq.~(\ref{Bi}) (orange, dashed).  }
       \label{integrand}
    \end{center}
  \end{figure}

\section{Edge-effect and theoretical description}\label{sec-SFA}

We consider ionization of an electron bound in an atomic potential  $V(r)$, in a  laser pulse with electric field $ \mathbf{E}(t)$. The asymptotic momentum distribution,
\begin{equation}\label{PMD}
  w(p)= |m(p)|^2,
\end{equation}
 is determined by  the SFA direct ionization amplitude \cite{Becker_2002}:
\begin{eqnarray}
m(p)=-i\int dt \langle\psi_p^V(t)|H_i(t)|\phi(t)\rangle,
\label{direct}
\end{eqnarray}
where $\phi(\mathbf{r},t)$ is the bound state wave function, $\psi^V_p(\mathbf{r},t)$ the Volkov wave function \cite{Volkov_1935}, $H_i(t)=\mathbf{r}\cdot \mathbf{E}(t)$ the electron interaction Hamiltonian with the laser field.
Atomic units are used throughout. The integrals in the  amplitudes of Eqs.~(\ref{direct}) are calculated in two ways, fully numerically and with the saddle-point approximation (SPA) analytically.



We illustrate the edge-effect on a 1D problem of ionization of an electron bound in a 1D zero-range potential  $V(x)=-\kappa\delta(x)$, in a half-cycle laser pulse with electric field $ E(t)=-E_0\cos^2(\omega t)$, with the field vanishing at $t<t_i$ and $t>t_f$. Here $\omega=0.05$ a.u., $\kappa=\sqrt{2I_p}=1$ a.u.,  $I_p$ is the ionization potential, $\gamma= \tilde{\omega} \kappa/E_0$ the Keldysh parameter,  with the effective frequency $\tilde{\omega}\equiv \sqrt{2}\omega$ related to the  $\cos^2$-pulse (the effective frequency is defined as $\tilde{\omega}=\sqrt{-E''(0)/E(0)}$ at the field maximum $t=0$).  We calculate PMD for different laser fields, using SFA amplitude of Eq.~(\ref{direct}), with the bound state wave function $\phi(x,t)=\sqrt{\kappa}\exp(-\kappa|x|+i\kappa^2/2t)$. The results are presented in Figs.~\ref{sin}-\ref{fig3}.

In strong fields, the PMD is a smooth function of the asymptotic momentum, see the case of $E_0=0.1$ a.u. for $\cos^2$-pulse in Fig.~\ref{sin}(a), and for  truncated-Gaussian pulses in Fig.~\ref{Gaussian}(a,c). In contrast, at weak fields PMD appears to be superimposed by the diffraction pattern due to the time-slit of the pulse edges, see  $E_0=0.05$ a.u. and $E_0=0.025$ a.u. in Fig.~\ref{sin}(b) and (c) for $\cos^2$-pulse, and Fig.~\ref{Gaussian}(b,d) for  Gaussian pulses, respectively (orange-dashed lines in figures correspond to SFA, and green-dotted lines to the TDSE numerical solutions). This effect is large in weak fields, when the ionization dynamical signal is weak, and strongly dependent on the pulse shape. In fact, in a Gaussian pulse $E(t)=-E_0 \exp[-(\omega t)^2]$ of the same effective frequency ($\omega=0.05$) as in $\cos^2$ one, the edge-effect gradually decreases with increasing of the Gaussian truncation. In particular, the edge-effect vanishes, i.e.  oscillations in PMD disappear, if a rather large truncation time   is applied, see the green-dotted lines in Figs.~\ref{Gaussian}(b,d), and~\ref{fig3} corresponding to $\omega (t_f-t_i)=16$. However, the edge-effect persists at smaller truncation time at the same field strength and the same frequency, see the dashed lines in Fig.~\ref{Gaussian}(b,d) corresponding to $\omega (t_f-t_i)=4$. While the use of Patchkovskii's smooth-truncated-Gaussian pulse of Ref.~\cite{Patchkovskii_2016} decreases the edge-effect, see red-dot-dashed line in Fig.~\ref{fig3}, however, at weak fields the boundary terms still contribute and contaminate the physical PMD.

In  Figs.~\ref{sin}-\ref{fig3}, we provide PMD via SFA, as well as via the numerical solution of TDSE. For strong fields $E_0=0.1$ and $E_0=0.05$, the SFA results are in close accordance with the numerical ones in any pulse. Deviations mainly originate from the Stark-shift that is not accounted in SFA, yielding slightly overestimated ionization probabilities. In weak fields, $E_0=0.025$ the results are still in accordance with long Gaussian pulses, but in short truncated-Gaussian pulses the edge effects are different in SFA and in TDSE, concurring only qualitatively.

\section{Conditions for the edge-effect appearance}\label{sec-conditions}

Generally, the smoother the switching-on and -off of the laser pulse, the less pronounced are the edge-effects. However, for a given smooth pulse shape there is a threshold intensity, below which the edge-effects again show up. This is illustrated for $\cos^n$-type pulses in Fig.~\ref{sinn}. We estimate the condition for the edge-effect appearance as follows. The switching-on/-off of the laser pulse results in appearance of high-energy components in the field spectrum. The edge-effect is induced by the high-energy component of the field with $\Omega \gtrsim I_p$, available in the spectrum of the pulse.
We characterize the edge-effect by the probability of photoionization via absorption of such a high-energy photon \cite{Delone_Krainov_book}:
\begin{equation}
\label{condition1}
  W_\Omega \sim \left(\frac{E_\Omega a}{\Omega}\right)^2,
\end{equation}
with the field strength $E_\Omega\sim E_0 (\omega/\Omega)^n$ of the high-frequency component $\Omega$, and the typical atomic length $a=1/\kappa$. The edge-effect will be visible if this probability is comparable with (or larger than) the strong-field ionization probability  \cite{Popov_2004} due to the monochromatic field of the effective frequency of the pulse:
\begin{equation}
\label{condition2}
  W_{SFI} \sim  \frac{E_a}{E_s}\exp\left\{ -\frac{2I_p}{\tilde{\omega}}\left[\left(1+\frac{1}{2\gamma^2}\right)\arcsinh \gamma-\frac{\sqrt{1+\gamma^2}}{2\gamma}\right]\right\}.
\end{equation}
where $E_s=E_0\sqrt{1+\gamma^2}$ is the field value at the time saddle-point.
Thus, the condition of the onset of the edge-effect is $W_\Omega\gtrsim W_{SFI}$. In Fig.~\ref{sinn} $W_\Omega=W_{SFI}$ corresponds to the  crossing point  of the red line representing $W_{SFI}$  with the corresponding one-photon probabilities $W_\Omega$ for different pulses. Thus, the edge-effect will be visible in the corresponding pulses with the field strength below the crossing points. The smoother the laser pulse, the smaller will be the laser intensity below which the edge-effect will emerge.
The edge-effect appears at low laser intensities, when the tunneling ionization signal is weak and becomes comparable with the diffraction signal. This effect hinders the understanding of nonadiabatic tunneling at large Keldysh parameters , as it conceals specific nonadiabatic features in PMD.

 \section{Separation of the edge-effect}\label{sec-U-contour}

In this section we put forward a method for separation of the edge effect and singling-out the dynamical features of PMD at given laser parameters. In the total PMD with the edge-effect, the dynamical signal is superimposed by the trivial diffraction pattern due to the time-slit of the pulse. Meanwhile, the dynamical signal is most interesting physically because it provides information on the nonadiabatic dynamics of the photoelectron in a weak field regime. As an  example we refer to structures inside the attoclock ring in a weak elliptically polarized laser field \cite{Ivanov_2014}, which also could be related to the unexplained large attoclock offset angles in the multiphoton regime \cite{Landsman_2014o}.

  \subsection{U-contour method}

Before introducing the method for separation of the edge-effects, let us to note that the edge-effect can be avoided in the calculation of PMD within SFA, when using SPA for the time-integration, e.g. see red-dashed line in Fig.~\ref{Gaussian}(b) \cite{Klaiber_2017}. However, in 3D cases and for a large range of PMD, e.g. in the attoclock, see Sec.~\ref{sec-attoclock}, it is a cumbersome procedure to find all saddle-points of the full PMD. Moreover, there still remains the question how to remove the edge-effect for TDSE. In the latter the only possibility is to use a Gaussian pulse with a very large truncation time, which requires extensive computational resources.

Here we propose a simple method for the calculation of the edge-effect-free PMD. The method mimics the saddle-point time-integration method for the ionization amplitude. In the first-order SFA the integrand $m(t)$ of the ionization amplitude
\begin{equation}
  m(p)=\int_{t_i}^{t_f} dt \,m(t),
  \label{m_p}
\end{equation}
has the form
\begin{eqnarray}
 m(t)={\cal C}\frac{(\mathbf{p}+\mathbf{A}(t))\cdot \mathbf{E}(t)}{\left[\left(\mathbf{p}+\mathbf{A}(t)\right)^2+\kappa^2\right]^2}\exp[-i S(t)+i\kappa^2/2 t],
 \label{m_t}
 \end{eqnarray}
with the classical action in the laser field $S(t)=\int^{t_f}_{t} ds (\mathbf{p}+\mathbf{A}(s))^2/2$, and the constant ${\cal C}_{1D}=-2 i \sqrt{\frac{2}{\pi }} \kappa ^{3/2}$ for the 1D case, and ${\cal C}_{3D}=1/(\sqrt{2\pi}\kappa){\cal C}_{1D}$ for the 3D case.
In the original Eq.~(\ref{m_p}) the time-integration runs along the real time-axis from the onset of the pulse $t_i$ to the end $t_f$, see Fig.~\ref{contours}. When one applies SPA, the original contour is deformed to the steepest-descent contour of the saddle-point.  Generally, one has to find all saddle-points corresponding to the given asymptotic momentum via an appropriate deformation of the initial contour of the time-integration. As is shown in Fig.~\ref{contours}, the integral along the steepest-descent contour equals to that along the U-contour (red in Fig.~\ref{contours}). Thus, the edge-free PMD can be obtained adding two integrals along the vertical contour to the main integral along the real axis (from $t_i$ to  $t_f$). This can be done analytically within SFA, as well as  numerically in TDSE solution.

 Further, we note that the U-contour can also be used to calculate half-cycle resolved ionization probabilities in long sinusoidal fields via truncating the field at the beginning and the end of the half-cycle of interest.

\subsection{Calculation of the time-integral along the vertical contour}

 \begin{figure*}
   \begin{center}
 \includegraphics[width=0.45\textwidth]{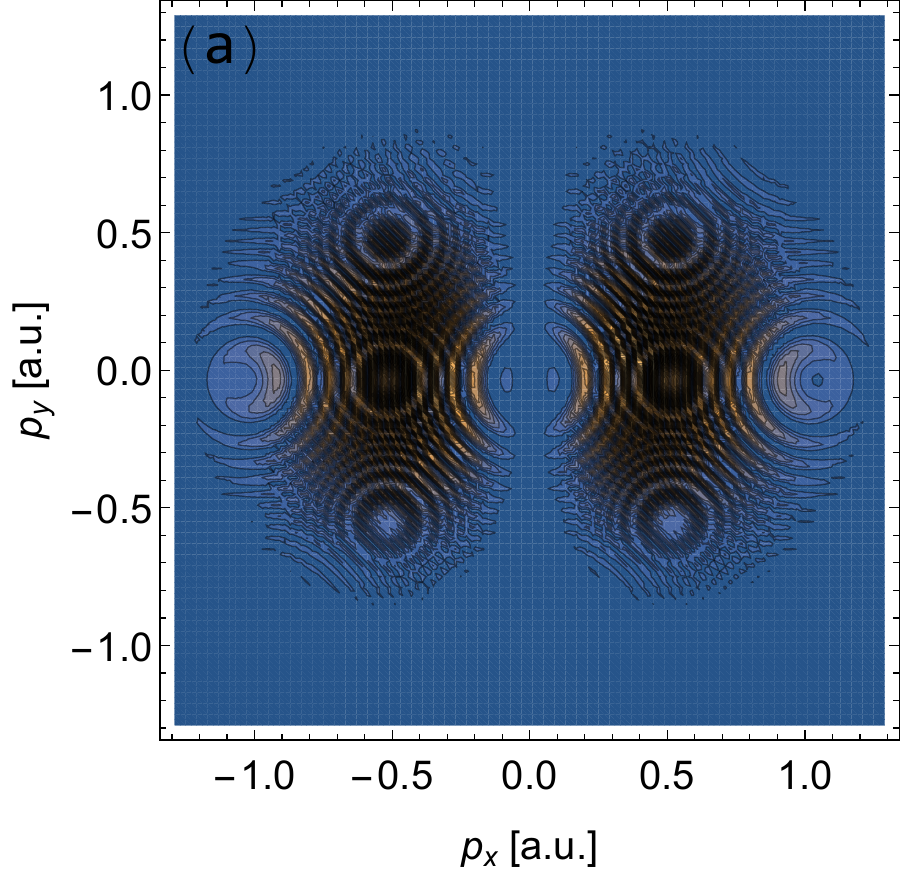}
\includegraphics[width=0.45\textwidth]{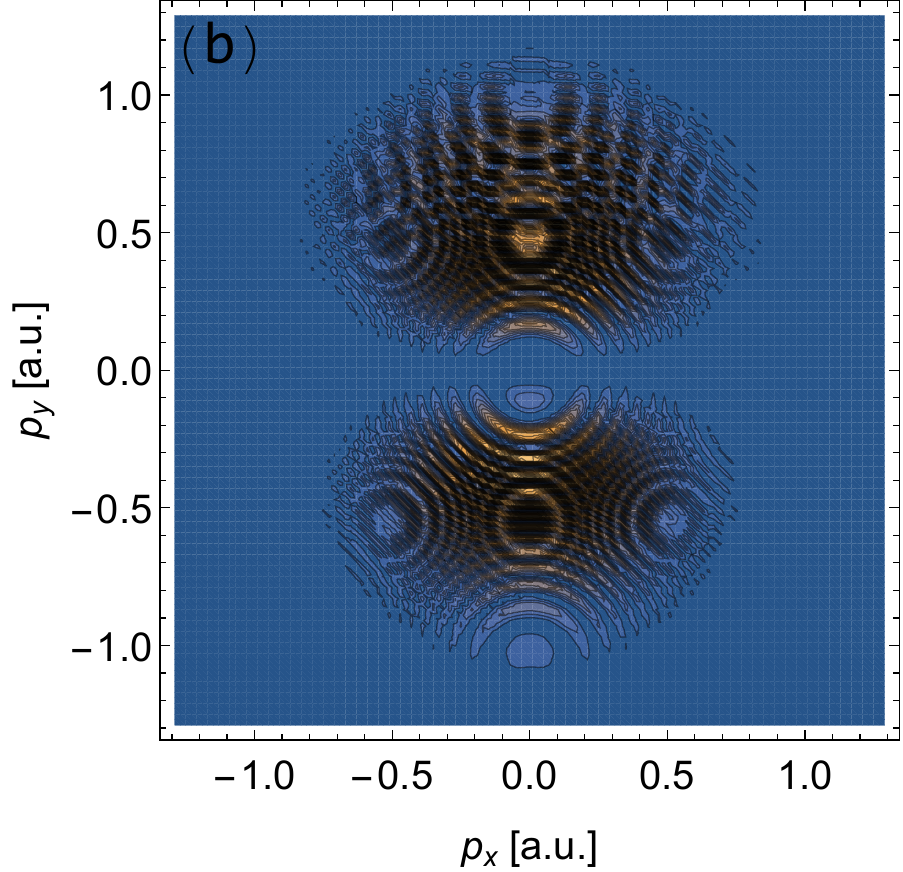}
           \caption{PMD in the attoclock via first-order SFA: (a) PMD with the edge effect included, (b)
                      PMD with the edge effect subtracted; The laser field is $\mathbf{E}(t)=-E_0\cos(\omega t/6)^2(\cos(\omega t),\epsilon\sin(\omega t))$, $\omega=0.062$ a.u., $\kappa=1.345$ a.u., $E_0=0.04 $ a.u., $\epsilon=0.87$ a.u. the field truncation is at $\omega t_i=-3\pi/\omega$ and $\omega t_f=3\pi/\omega$. }
       \label{elliptical}
    \end{center}
  \end{figure*}

The calculation of  the time-integrals along the vertical contours in Fig.~\ref{contours}, $ C_i$ and $C_f$,
is facilitated by the fact that the integrand is exponentially suppressed at large imaginary times, see Fig.~\ref{integrand}, and only the beginning of the contour close to the real axis gives the main contribution to the integral.
 In  order to account for the edge-effect analytically, we approximate the prefactor and the exponential of the integrand $m(t)$ near the truncation points $t_i$ and $t_f$:
\begin{eqnarray}
m_{C_i}(t)&\approx &{\cal C}\sum_{n}\frac{(p+A(t_i))A^{(n+1)}(t_i)(t-t_i)^n}{n!\left((p+A(t_i))^2+\kappa^2\right)^2} \nonumber \\
&\times & \exp\left[-i S(t_i)+i \kappa^2/2t_i\right.\nonumber\\
&+& \left.i(\kappa^2/2+(p+A(t_i))^2/2) (t-t_i)\right]\\
m_{C_f}(t) & \approx &{\cal C} \sum_{n}\frac{(p+A(t_f))A^{(n+1)}(t_f)(t-t_f)^n}{n!\left((p+A(t_f))^2+\kappa^2\right)^2}\nonumber\\
&\times & \exp\left[-i S(t_f)+i\kappa^2/2t_f\right. \nonumber\\
&+& \left.i(\kappa^2/2+(p+A(t_f))^2/2) (t-t_f)\right],
\end{eqnarray}
where the summation over $n$
begins from the first non-vanishing derivative of the function $A'(t)$ up to the next third orders at weak fields and terms of the order of $A'(t_i)^2$ or $A'(t_f)^2$ are neglected.
These approximated integrands can now be integrated along the steepest descent contour at the truncation points. Since $\kappa^2/2+(p+A(t_i))^2/2$ and $\kappa^2/2+(p+A(t_f))^2/2$ are real numbers, the contour is vertically aligned in the complex plane starting at $t_i$ or $t_f$, respectively. The integration yields
\begin{eqnarray}
m_{C_i}(p)&=&{\cal C} \sum_{n}2^{n+1}\exp[-i S(t_i)+i\kappa^2/2t_i) ]\nonumber\\
&\times &\frac{(p+A(t_i))A^{n+1}(t_i)}{((p+A(t_i))+\kappa^2)^{n+1}} \\ \label{Bi}
m_{C_f}(p)& =&{\cal C}\sum_{n} 2^{n+1}\exp[-i S(t_f)+i\kappa^2/2t_f) ]\nonumber\\
&\times& \frac{(p+A(t_f))A^{n+1}(t_f)}{((p+A(t_f))+\kappa^2)^{n+1}}. \label{Bf}
\end{eqnarray}
The approximated integrand function of Eq.~(\ref{Bi}) is shown in Fig.~\ref{integrand}. It coincides with the analytical one. Thus, using expressions of Eqs.~(\ref{Bi}), (\ref{Bf}) the contribution of the vertical contours $C_i$, $C_f$ can be subtracted analytically, which corresponds to the subtraction of the edge-effect.

\subsection{Edge-effect subtraction in the numerical solution of TDSE}

The PMD according to numerical solution of TDSE can be written as:
\begin{equation}
	\label{eq:pmd_num}
    w(p) = |m(p)|^2 = |\langle{\psi_p^V(x, t_f)}| U(t_f, t_i) |{\phi(x, t_i)}\rangle|^2,
\end{equation}
where the time evolution operator $U(t_f, t_i)$ can be obtained through the normal Schr\"{o}dinger equation
\begin{equation}
	\label{eq:u_sch}
    U(t_f, t_i)=\mathcal{T} \exp\left[ -i \int_{t_i}^{t_f} dt \, H(t) \right],
\end{equation}
with $H(t)=\hat{\mathbf{p}}^2/2m + \mathbf{r} \cdot \mathbf{E}(t) + V(\mathbf{r})$ and $\mathcal{T}$ being the time ordering operator, or based on the Dyson equation
\begin{equation}
	\label{eq:u_dys}
    U(t_f, t_i)=U_0(t_f, t_i) - i \int_{t_i}^{t_f}dt \, U(t_f, t)H_i(t)U_0(t, t_i),
\end{equation}
with $H_i(t) = \mathbf{r} \cdot \mathbf{E}(t)$ being the interaction Hamiltonian and $U_0(t_2, t_1)$ being the field free time evolution operator.

The ionization amplitude $m(p)$ along the real axis is calculated as
\begin{equation}
	\label{eq:mp_real}
    m_{C_r}(p) = \langle{\psi_p^V(t_f)}| \mathcal{T} \exp\left[ -i \int_{t_i}^{t_f} dt \, H(t) \right]|  {\phi(t_i)}\rangle
\end{equation}
employing the traditional time-splitting operator method. Along the vertical contours ($C_i$ and $C_f$), on the other hand, the amplitude is obtained through
\begin{eqnarray}
	\label{eq:mp_aimag}
    m_{C_i}(p) = - i \int_{C_i}dt \,\langle \psi_p^V(t_f) |U(t_f, t)H_i(t)U_0(t, t_i) | \phi(t_i)\rangle, \\
    m_{C_f}(p) = - i \int_{C_f}dt \, \langle \psi_p^V(t_f)| U(t_f, t)H_i(t)U_0(t, t_i) |\phi(t_i)\rangle.
\end{eqnarray}
Here $U(t_f, t)$ is the exact time evolution operator in the numerical simulation rather than the time evolution operator in the laser pulse only, which is used in the SFA calculations.

Finally, the total PMD is calculated as
\begin{equation}
	\label{eq:pmd_num_fin}
    w(p) = |m_{C_r}(p)+m_{C_i}(p)+m_{C_f}(p)|^2 .
\end{equation}

\section{Attoclock}\label{sec-attoclock}

The proposed U-contour method proves very efficient for the calculation of PMD of the attoclock at weak laser intensities. We calculate PMD via 3D first-order SFA in the attoclock case with an elliptically polarized laser field
\begin{equation}\label{EEE}
  \mathbf{E}(t)=-E_0\cos(\omega t/6)^2\left[\cos(\omega t),\epsilon\sin(\omega t)\right],
\end{equation}
with $\omega=0.062$ a.u., $\kappa=1.345$ a.u., $E_0=0.04 $ a.u., $\epsilon=0.87$ a.u. and truncation at $\omega t_i=-3\pi/\omega$ and $\omega t_f=3\pi/\omega$. The time integral in Eqs.~(\ref{m_p})-(\ref{m_t}) is calculated numerically  and the edge terms are subtracted analytically as shown above.

The corresponding momentum distribution is shown in Fig.~\ref{elliptical}. Note that in a long Gaussian pulse the PMD coincides with the edge-effect-free result of the U-contour method. One can see that the edge-effects significantly disturb PMD, even changing the topology of the distribution. Usually the Coulomb field effect increases the ionization signal, and makes the edge-effect relevant at smaller field strength than in the short-range potential.

\section{Conclusion}\label{sec-Conclusion}

We have developed a new method to remove the edge-effect of the laser pulse because of the diffraction from the time slit created by the edges, in PMD of tunnel-ionized electrons. The method consists of replacing the original time-integral in the ionization amplitude along the real time axis with the, so-called, U-contour, adding two integrals along the imaginary time axis,  starting at the time edges of the laser pulse. The method can be applied  analytically for SFA, as well as for the numerical solution of TDSE.  The edge-effect adds a trivial diffraction patterns originated from the time-edges of the laser pulse, which disappear when using more smooth laser pulses (long-truncated-Gaussian pulse) of the same frequency and intensity. The edge-effect hides the physical structures in PMD due to nonadiabatic processes in weak laser fields that underlines the important application of the proposed method to reveal  the dynamical signal of strong-field ionization in the deep nonadiabatic regime.


\bibliography{strong_fields_bibliography}

\end{document}